\documentclass[journal,10pt]{IEEEtran} 

\usepackage[T1]{fontenc}

\usepackage{graphicx}
\usepackage[export]{adjustbox}
\usepackage[dvipsnames]{xcolor}
\usepackage{bbm}

\graphicspath{{./Figs/}}
\usepackage{subfigure}
\usepackage{stfloats}
\ifCLASSINFOpdf
\else
\fi

\def\T{\mathrm{T}}
\def\L{\mathrm{L}}
\def\Fn{\mathrm{F}_n}
\def\F{\mathrm{F}}
\def\N{N}
\def\tc{\theta_{\mathrm{C}}}
\def\tmin{\theta_{\mathrm{min}}}
\def\Nl{N_{\mathrm{max}}^{\mathrm{lin}}}
\def\Nc{N_{\mathrm{max}}^{\mathrm{cir}}}
\def\re{r_\mathrm{e}}
\def\gL{\gamma_{\L}}
\def\gFn{\gamma_{\Fn}}
\def\PL{P_{\L}}
\def\PFn{P_{\Fn}}
\def\g{\gamma}
\def\gth{\g_{\mathrm{th}}}
\def\Pout{P_{\mathrm{out}}}

\usepackage{amssymb}
\usepackage{amsmath}
\usepackage[cmintegrals]{newtxmath}
\usepackage{stackengine}

\usepackage{algorithmic}
\usepackage{algorithm}

\usepackage{amsthm}
\usepackage{xpatch}

\newtheorem{rem}{Remark}

\xpatchcmd{\proof}{\hskip\labelsep}{\hskip5\labelsep}{}{}

\usepackage{mathtools}

\DeclarePairedDelimiter{\floor}{\lfloor}{\rfloor}

\usepackage{array}

\begin{document}
\title{Satellite Clusters Flying in Formation: Orbital Configuration-Dependent Performance Analyses}

\author{Dong-Hyun Jung, Joon-Gyu Ryu, and Junil Choi\\

\thanks{D.-H. Jung and J.-G. Ryu are with the Satellite Communication Research Division, Electronics and Telecommunications Research Institute, Daejeon, South Korea (e-mail: \{dhjung, jgryurt\}@etri.re.kr).}
\thanks{J. Choi is with the School of Electrical Engineering, KAIST, Daejeon, South Korea (e-mail: junil@kaist.ac.kr).}
\vspace{-1cm}
}
\maketitle

\begin{abstract}

This paper considers a downlink satellite communication system where a \textit{satellite cluster}, i.e., a satellite swarm consisting of one leader and multiple follower satellites, serves a ground terminal. The satellites in the cluster form either a linear or circular formation moving in a group and cooperatively send their signals by maximum ratio transmission precoding. We first conduct a coordinate transformation to effectively capture the relative positions of satellites in the cluster. Next, we derive an exact expression for the orbital configuration-dependent outage probability under the Nakagami fading by using the distribution of the sum of independent Gamma random variables. In addition, we obtain a simpler approximated expression for the outage probability with the help of second-order moment-matching. We also analyze asymptotic behavior in the high signal-to-noise ratio regime and the diversity order of the outage performance. Finally, we verify the analytical results through Monte Carlo simulations. Our analytical results provide the performance of satellite cluster-based communication systems based on specific orbital configurations, which can be used to design reliable satellite clusters in terms of cluster size, formation, and orbits.

\textbf{\emph{Index terms}} --- Satellite communication systems, satellite clusters, outage probability, orbital configuration.
\\
\end{abstract}

\IEEEpeerreviewmaketitle

\vspace{-0.6cm}

\section{Introduction}\label{Sec:Intro_general}
\IEEEPARstart{S}{atellite} communications have been considered as a viable option for providing global Internet services, thanks to the extensive coverage of satellites. To achieve this, non-terrestrial networks (NTNs) have been developed and standardized by the 3rd-Generation Partnership Project (3GPP) as part of the fifth-generation standard since Release 15 [\ref{Ref:3GPP_38.811}], [\ref{Ref:3GPP_38.821}]. NTNs consist of flying objects, such as satellites, unmanned aerial vehicles, and high-altitude platforms, that can be combined with terrestrial networks to offer communication services to ground terminals and aerial vehicles such as drones, airplanes, and urban air mobility vehicles. Several companies such as SpaceX, OneWeb, and Telesat have plans to launch multiple low Earth orbit (LEO) satellite constellations in the near future to boost system throughput [\ref{Ref:Pachler}]. 

Once a considerable number of satellites have been launched, sending additional satellites into orbit could cause inter-satellite interference, which would likely result in only a minor increase in network throughput. Thus, recent research has focused on the utilization of satellite clusters, i.e., clusters of small satellites located in close proximity to one another, as a potential means to improve network performance through collaboration [\ref{Ref:Liu}]-[\ref{Ref:Popov}].
Various architectures and methods for small satellites' formation flying were introduced in [\ref{Ref:Liu}] including the leader-following method.
System-level performance of satellite cluster-based communication systems was evaluated in [\ref{Ref:Jung}] where 3GPP-based network architectures for satellite clusters were also proposed.
Satellite-based distributed multi-input multi-output transmissions were considered in [\ref{Ref:Barton}] for deep-space communications where the advantages of distributed beamforming among multiple satellites were demonstrated in terms of the trade-off between spectral and energy efficiencies.
Virtual beamforming for satellite clusters was investigated in [\ref{Ref:Yu}], and output feedback control algorithms for satellite formation flying were proposed and demonstrated in [\ref{Ref:Popov}] to construct the formations with low energy costs and robustness to perturbations.

Although the previous work [\ref{Ref:Jung}] has evaluated the system-level performance showing the superiority of satellite clusters to conventional constellations, few studies have analyzed the performance of satellite clusters considering specific flying formations and orbital configuration. The scalability of a cluster, i.e., the number of satellites in the cluster, may vary depending on flying formations due to different spatial efficiencies. 
Moreover, the orbital elements, such as the inclination, the argument of latitude, and the right ascension of the ascending node (RAAN), are the key parameters that have a significant impact on communication performance.
Motivated by this, we aim to analyze the orbital configuration-dependent performance of satellite cluster-based communication systems where the satellites in the cluster fly in either a linear or circular formation. We characterize the distance between the terminal and the satellites in the cluster by conducting a coordinate transformation. Using these distance characteristics, we obtain both exact and approximated expressions for the outage probability. Furthermore, we analyze the asymptotic behavior of the outage performance in terms of the diversity order. Finally, we verify the derived expressions using Monte Carlo simulations.

\vspace{-0.2cm}
\section{System Model}\label{Sec:System_model}
\begin{figure*}
\begin{center}
\includegraphics[width=2\columnwidth]{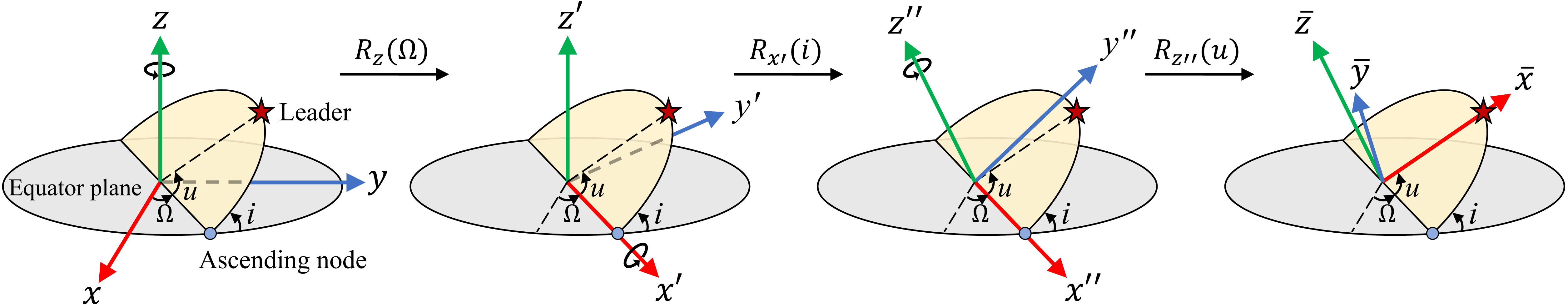}
\end{center}
\setlength\abovecaptionskip{.25ex plus .125ex minus .125ex}
\setlength\belowcaptionskip{.25ex plus .125ex minus .125ex}
\caption{Procedure of the coordinate transformation $\mathcal{C} \rightarrow \mathcal{C}^{\prime}$ with the sequence of axes $z-x^{\prime}-z^{\prime\prime}$. This transformation locates the position of the leader on the $\bar{x}-$axis in the new coordinate frame $\mathcal{C}^{\prime}$.}
\vspace{-10pt}
\label{Fig:coordi_transform}
\end{figure*}

We consider a downlink satellite communication system where a satellite cluster consisting of one leader satellite $\L$ and $\N$ follower satellites $\Fn$, $n\in\{1,2,\cdots,\N\}$, serves a terminal $\T$.
It is assumed that all the satellites in the cluster have the same altitudes of $a$.
The leader moves along a circular orbit, which we call a reference orbit of the satellite cluster, and is configured by the inclination $i$, the argument of latitude $u$, and the RAAN $\Omega$ (the meaning of these parameters will become clear in Fig. \ref{Fig:coordi_transform}).
The orbits of the followers are configured based on the reference orbit so that the satellites in the cluster move in a group. 

\begin{figure}
\begin{center}
\includegraphics[width=\columnwidth]{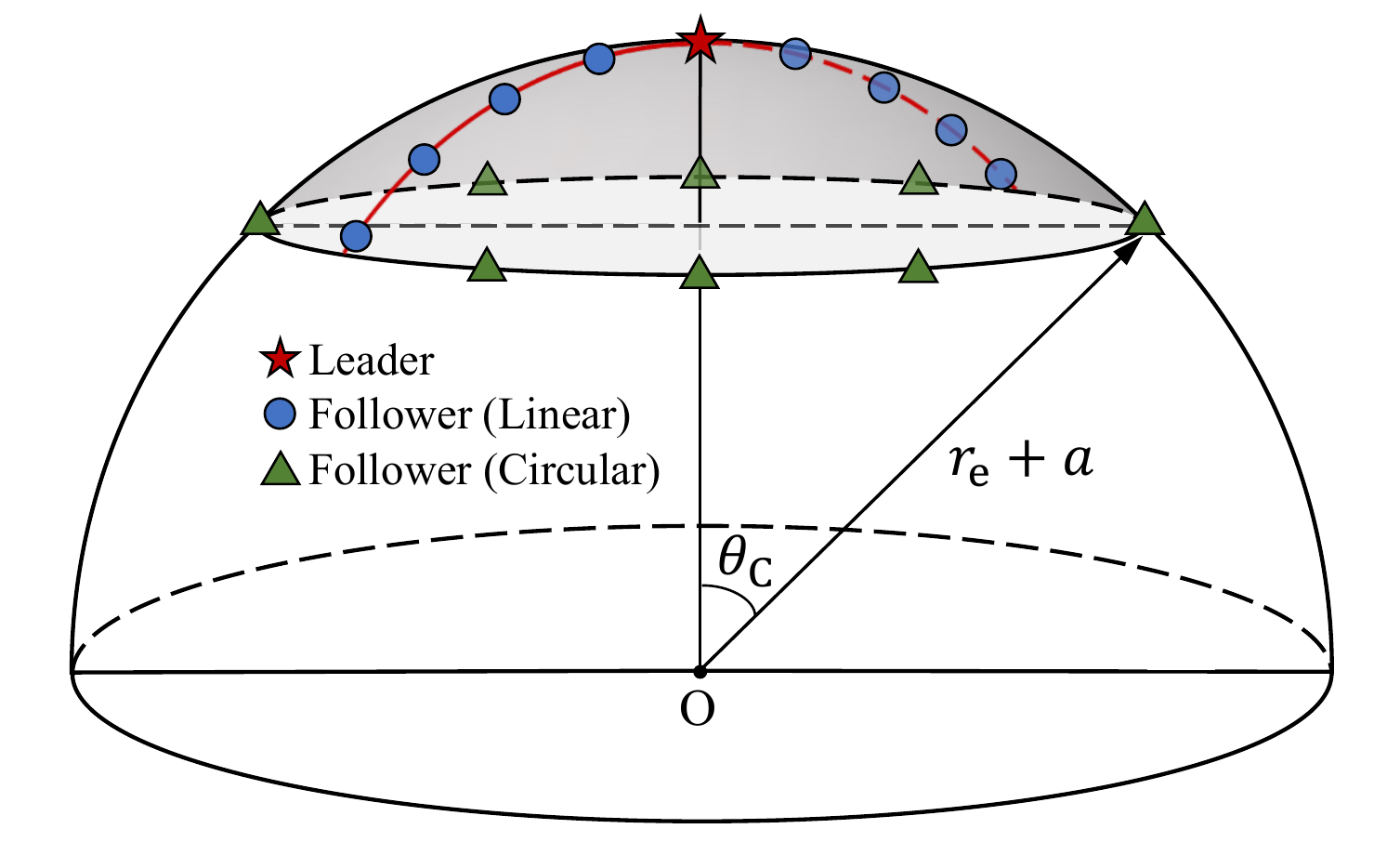}
\end{center}
\setlength\abovecaptionskip{.25ex plus .125ex minus .125ex}
\setlength\belowcaptionskip{.25ex plus .125ex minus .125ex}
\caption{An example of linear and circular clusters with one leader and eight followers where the red line is the reference orbit of the cluster.}
\vspace{-10pt}
\label{Fig:linear_vs_circular}
\end{figure}

Two types of formation flying are considered: linear and circular clusters as shown in Fig. \ref{Fig:linear_vs_circular} where $\mathrm{O}$ is the Earth's center, $\re$ is the Earth's radius, and $\tc$ is the polar angle of the spherical cap, i.e., the area where the satellites in a cluster can be located.
The linear cluster is a cluster in which the followers are located in the same orbit as the leader’s (the reference orbit) keeping a close distance to adjacent satellites. 
In the circular cluster, the followers have slightly different orbital configurations to keep a circular formation when viewed from the Earth and are equidistant from the leader [\ref{Ref:Popov}].
Assume that the satellites in the cluster should be apart from one another at a distance larger than $d_0$ due to limited position control accuracy and collision risks.
From this assumption, given $\tc$, we can obtain the maximum number of followers that can be deployed in the linear or circular cluster as $\Nl = 2 \bigl\lfloor{ \frac{\tc}{\tmin} \bigr\rfloor}$ or $\Nc = \bigl\lfloor{\frac{2\pi\sin\tc}{\tmin}\bigr\rfloor}$
where $\floor{\cdot}$ is the floor function, and $\tmin=2 \sin^{-1}\left(\frac{d_0}{2(\re+a)}\right)$ is the minimum polar angle between two adjacent satellites.
The proof of these results is given in Appendix.

The channel coefficient for the link between the terminal and the satellite $i\in\{\L,\F_1,\F_2,\cdots,\F_\N\}$, denoted by $h_i$, is expressed as $h_i=\tilde{h}_i\sqrt{\ell_i}$ where $\tilde{h}_i$ is the small-scale fading coefficient, and $\ell_i$ is the path loss between the satellite $i$ and the terminal.
For small-scale fading, we use the Nakagami-$m$ distribution which has been well-adopted for satellite channels due to its versatility of modeling [\ref{Ref:Park}]. For example, when $m=1$ and $m=\frac{(K+1)^2}{2K+1}$, the Nakagami-$m$ distribution becomes the Rayleigh and Rician-$K$ distributions, respectively.
Note that the channel gain of the Nakagami fading model follows the Gamma distribution, i.e., $|\tilde{h}_i|^2\sim\Gamma(m_i,\frac{1}{m_i})$, of which the CDF is given by $F_{|\tilde{h}_i|^2}(x)=\frac{\gamma(m_i,m_i x)}{\Gamma(m_i)}$ where $\gamma(a,x)=\int_0^x t^{a-1} e^{-t} dt$ is the lower incomplete Gamma function, and $\Gamma(x)$ is the Gamma function.
The path loss is given as $\ell_i = \left(\frac{c}{4\pi f_{\mathrm{c}}}\right)^2 d_{i}^{-\alpha_i}$ where $c$ is the speed of light, $f_{\mathrm{c}}$ is the carrier frequency, $d_i$ is the link distance, and $\alpha_i$ is the path loss exponent.

Let $\mathbf{h}$ denote the vector of channel coefficients for the cluster, i.e., $\mathbf{h} = [h_{\L},h_{\F_1},h_{\F_2},\cdots,h_{\F_{\N}}]^T$. We assume that the satellite cluster adopts the maximum ratio transmission (MRT) where the beamforming vector is given by $\frac{\mathbf{h}}{||\mathbf{h}||}$. When the MRT is used, the received signal-to-noise ratio (SNR) at the terminal is given by
\begin{align}
\gamma 
    \!=\! \gL + \sum_{n=1}^{\N} \gFn
    \!=\! \frac{\PL G_{\mathrm{t,L}} G_{\mathrm{r}} |h_{\L}|^2}{N_0 W} \!+\! \sum_{n=1}^{\N} \!\frac{\PFn G_{\mathrm{t},\mathrm{F}_n} G_{\mathrm{r}} |h_{\Fn}|^2}{N_0 W}
\end{align}
where $P_i$ is the transmit power of the satellite $i$, $G_{\mathrm{t},i}$ is the transmit antenna gain, $G_{\mathrm{r}}$ is the receive antenna gain, $N_0$ is the noise power spectral density, and $W$ is the bandwidth.

\section{Performance Analysis}
In this section, we first conduct the coordinate transformation to effectively capture the relative positions of satellites in the cluster. Next, we characterize the distance between the satellites in the cluster and the terminal by using the relative positions of the satellites. Then, we derive an exact expression for the outage probability of the system. In addition, we obtain an approximated outage probability expression using the moment-matching method and asymptotic behavior in the high SNR regime.

\subsection{Coordinate Transformation}
We assume the Earth-centered coordinate frame $\mathcal{C}$, i.e., the center of the Earth is the origin.
Let $\mathrm{P}_{\mathrm{X}}^{\mathcal{A}}$ denote the position of a point $\mathrm{X}$ with respect to a coordinate frame $\mathcal{A}$. Then, the position of the Earth's center $\mathrm{O}$ is expressed as $\mathrm{P}_{\mathrm{O}}^{\mathcal{C}}=(0,0,0)$.
Since the followers are located based on the leader's position, we adopt a coordinate transformation using the \textit{proper Eular angles}\footnote{The Eular angles are used to describe the orientation of a spacecraft with respect to a fixed coordinate system. Three successive intrinsic rotations of axes, whose magnitudes are the Euler angles, can reach any target orientation. The proper Euler angles use the same axis for both the first and third elemental rotations, e.g., $z-x^{\prime}-z^{\prime\prime}$ or $x-y^{\prime}-x^{\prime\prime}$.} with the sequence of rotation axes $z-x^{\prime}-z^{\prime\prime}$ as described in Fig. \ref{Fig:coordi_transform}.
This coordinate transformation efficiently characterizes the relative positions of satellites in the cluster and includes the rotations about the three axes by the angles equal to the RAAN $\Omega$, the inclination $i$, and the argument of latitude $u$, respectively.
Let $R_x(\phi)$, $R_y(\phi)$, and $R_z(\phi)$ be the basic three-dimensional rotation matrices about the $x-$, $y-$, and $z-$axes, i.e.,
\begin{align}
    R_x(\phi) = \begin{bmatrix}
                1 & 0 & 0\\
                0 & \cos\phi & -\sin\phi\\
                0 & \sin\phi & \cos\phi
                \end{bmatrix},
\end{align}
\begin{align}
    R_y(\phi) = \begin{bmatrix}
                \cos\phi & 0 & \sin\phi\\
                0 & 1 & 0\\
                -\sin\phi & 0 & \cos\phi
                \end{bmatrix},
\end{align}
\begin{align}
    R_z(\phi) = \begin{bmatrix}
                \cos\phi & -\sin\phi & 0\\
                \sin\phi & \cos\phi & 0\\
                0 & 0 & 1
                \end{bmatrix}.
\end{align}
Then, the intrinsic rotations with the sequence of axes $z-x^{\prime}-z^{\prime\prime}$ correspond to the rotation matrix $R=R_z(\Omega)R_{x}(i)R_{z}(u)$ 
and give the new coordinate frame $\mathcal{C^{\prime}}$ with $\bar{x}-$, $\bar{y}-$, and $\bar{z}-$axes.

\begin{rem}\label{Rem:pos_leader}
After the coordinate transformation with the sequence of rotation axes $z-x^{\prime}-z^{\prime\prime}$, the reference orbital plane lies on $\bar{x}\bar{y}$ plane where the leader is located on the $\bar{x}-$axis, i.e., $\mathrm{P}_{\L}^{\mathcal{C}^{\prime}}=(\re+a,0,0)$.
\end{rem}
\begin{rem}\label{Rem:pos_terminal}
Let $\mathcal{A}$ and $\mathcal{B}$ denote two different coordinate frames, and $R^{\mathcal{AB}}$ be the rotation matrix that transforms the frame $\mathcal{A}$ into the frame $\mathcal{B}$. 
Then, the position of an arbitrary point $\mathrm{X}$ in the frame $\mathcal{A}$ can be re-expressed in the new coordinate frame $\mathcal{B}$ as $\mathrm{P}_{\mathrm{X}}^{\mathcal{B}}=(R^{\mathcal{AB}})^{-1} \mathrm{P}_{\mathrm{X}}^{\mathcal{A}}$.
With the fact that the inverse matrix of a rotation matrix $R(\phi)$ is $R^{-1}(\phi)=R(-\phi)$, the position of the terminal in the new coordinate frame $\mathcal{C}^{\prime}$ is given by
$ \mathrm{P}_{\T}^{\mathcal{C}^{\prime}} = R^{-1} \mathrm{P}_{\T}^{\mathcal{C}} = R_{z}(-u)R_{x}(-i)R_z(-\Omega) \mathrm{P}_{\T}^{\mathcal{C}}$.
\end{rem}
For notational simplicity, we drop the superscript $\mathcal{C}^{\prime}$ from $\mathrm{P}_{\mathrm{X}}^{\mathcal{C}^{\prime}}$, i.e., $\mathrm{P}_{\mathrm{X}}^{\mathcal{C}^{\prime}} \rightarrow \mathrm{P}_{\mathrm{X}}$, because we only deal with the positions in the new coordinate frame $\mathcal{C}^{\prime}$ hereafter.

\subsection{Distance Characteristics}\label{Sec:dist_ch}
In the linear cluster, the followers are lined up along the leader's orbit as shown in Fig. \ref{Fig:linear_vs_circular}. Thanks to the coordinate transformation, the leader's orbit is on the $\bar{x}\bar{y}$ plane. Hence, the positions of the followers can be readily obtained by rotating the leader's position about the $\bar{z}-$axis where the angle of rotation for the follower $\Fn$ is given by
\begin{align}
\Delta\theta_n = &
    \begin{cases}
        \tc-\frac{2\tc}{\N}(n-1),& \mbox{if } n \le \frac{\N}{2},\\ 
        \tc-\frac{2\tc}{\N}n, & \mbox{otherwise.}
     \end{cases}
\end{align}
The position of the follower $\F_n$ is then given by
\begin{align}\label{eq:pos_follower_lin}
&\mathrm{P}_{\Fn}
    = R_{\bar{z}}(\Delta\theta_n)\mathrm{P}_{\L}\nonumber\\
    &= \begin{bmatrix}
        \cos{\Delta\theta_n} & -\sin{\Delta\theta_n} & 0\\
        \sin{\Delta\theta_n} & \cos{\Delta\theta_n} & 0\\
        0 & 0 & 1
      \end{bmatrix}
      \begin{bmatrix}
        \re+a \\
        0 \\
        0
      \end{bmatrix}
    =\begin{bmatrix}
        (\re+a)\cos{\Delta\theta_n} \\
        (\re+a)\sin{\Delta\theta_n} \\
        0
      \end{bmatrix}.
\end{align}

\begin{figure*}[b]
\setcounter{equation}{6}
\normalsize \hrulefill \vspace*{4pt}
\begin{align}\label{eq:pos_follower_cir}
    \mathrm{P}_{\Fn}
    &= R_{\bar{x}}\left(\beta_0+\frac{2\pi n}{\N}\right) R_{\bar{y}}(\tc) \mathrm{P}_{\L}\nonumber\\
    &= \begin{bmatrix}
            1 & 0 & 0\\
            0 & \cos{\left(\beta_0+\frac{2\pi n}{\N}\right)} & -\sin{\left(\beta_0+\frac{2\pi n}{\N}\right)}\\
            0 & \sin{\left(\beta_0+\frac{2\pi n}{\N}\right)} & \cos{\left(\beta_0+\frac{2\pi n}{\N}\right)}
        \end{bmatrix}
        \begin{bmatrix}
                \cos{\tc} & 0 & \sin{\tc}\\
                0 & 1 & 0\\
                -\sin{\tc} & 0 & \cos{\tc}
        \end{bmatrix}
      \begin{bmatrix}
        \re+a \\
        0 \\
        0
      \end{bmatrix}
      =\begin{bmatrix}
        (\re+a)\cos{\tc} \\
        (\re+a)\sin{\left(\beta_0+\frac{2\pi n}{\N}\right)}\sin{\tc} \\
        (\re+a)\cos{\left(\beta_0+\frac{2\pi n}{\N}\right)}\sin{\tc}
      \end{bmatrix}
\end{align}
\end{figure*}

In the circular cluster, the followers are at the circular edge of the spherical cap as shown in Fig. \ref{Fig:linear_vs_circular}. The positions of the followers $\Fn$ can be characterized with two successive extrinsic rotations: i) Rotation about the $\bar{y}-$axis by $\tc$ and ii) rotation about the $\bar{x}-$axis by $\beta_0+\frac{2\pi n}{\N}$. The corresponding rotation matrices for the two rotations are given by $R_{\bar{y}}(\tc)$ and $R_{\bar{x}}\left(\beta_0+\frac{2\pi n}{\N}\right)$, respectively.
The position of the follower $\F_n$ is given by \eqref{eq:pos_follower_cir} at the bottom of this page. Using the position of the terminal given in Remark \ref{Rem:pos_terminal} and the positions of the followers in \eqref{eq:pos_follower_lin} and \eqref{eq:pos_follower_cir}, the distances between the terminal and the followers in the cluster can be readily obtained, but the derivation is omitted due to the space limitation. These distances will be used to calculate the path loss in the next subsection.
\setcounter{equation}{7}

\subsection{Outage Probability}
An outage occurs at the terminal when the received SNR falls below an SNR threshold $\gth$.
Mathematically, the outage probability of the system is given by 
\begin{align}\label{eq:Pout_first}
\Pout
    = \mathbb{P}[\g < \gth]
    = \mathbb{P}\left[\sum_{i}a_i H_i < \gth\right]
\end{align}
where $a_i=\frac{P_i G_{\mathrm{t},i} G_{\mathrm{r}} \ell_i}{N_0 W}$ and $H_i=|\tilde{h}_i|^2$. To further derive the expression for the outage probability, we first let $Y_i = a_i H_i$ and obtain the distribution of $Y_i$. The CDF of $Y_i$, $i=\{\L,\F_1,\F_2,\cdots,\F_{\N}\}$, is given by
\begin{align}
F_{Y_i}(x)
    = \mathbb{P}[Y_i < x]
    = \mathbb{P}\left[H_i < \frac{x}{a_i}\right]
    = \frac{\g\left(m_i, \frac{m_i x}{a_i}\right)}{\Gamma(m_i)}.
\end{align}
Note that the random variable $Y_i$ is still a scaled-Gamma random variable, which follows the Gamma distribution with the shape and rate parameters $m_i$ and $a_i/m_i$, respectively, i.e., $Y_i \sim \Gamma\left(m_i, \frac{a_i}{m_i}\right)$.

To obtain an exact expression for the outage probability, we need the distribution of the sum of independent Gamma random variables with different shape and scale parameters. 
With the help of the result in [\ref{Ref:Moschopoulos}] providing the closed-form expression for this distribution, the PDF of $\sum_i Y_i$ is given by
\begin{align}
f_{\sum_i Y_i}(x) = L\sum_{l=0}^{\infty} \frac{\delta_{l} x^{\rho+l-1} e^{-\frac{x}{q}}}{\Gamma(\rho+l)q^{\rho+l}}
\end{align}
where $L = \prod_i \left(q m_i/a_i\right)^{m_i}$, $\rho = \sum_i m_i$, $q=\min_i \left(a_i/m_i\right)$, and $\delta_l$ can be obtained by recursion from 
$\delta_l = \frac{1}{l}\sum_{j=1}^{l} j g_j \delta_{l-j}$, 
$g_j = \sum_i \frac{m_i}{j} \left(1-\frac{q m_i}{a_i}\right)^j$,
and $\delta_0 = 1$.
Using this PDF, the outage probability can be further derived as
\begin{align}\label{eq:Pout_exact_fin}
&\Pout
    = \int_{0}^{\gth} f_{\sum_i Y_i}(x) dx \nonumber\\
    &= L\sum_{l=0}^{\infty} \frac{\delta_{l} }{\Gamma(\rho+l)q^{\rho+l}} \int_{0}^{\gth} x^{\rho+l-1} e^{-\frac{x}{q}} dx \nonumber\\
    &\mathop=^{(a)} L\sum_{l=0}^{\infty} \frac{\delta_{l} }{\Gamma(\rho+l)} \int_{0}^{\gth/q}  t^{\rho+l-1} e^{-t} dt \mathop=^{(b)} L\sum_{l=0}^{\infty} \frac{\delta_{l}\g(\rho + l, \frac{\gth}{q}) }{\Gamma(\rho+l)}
\end{align}
where ($a$) follows from the change of variables $t=x/q$, and ($b$) follows from the definition of the lower incomplete Gamma function. The expression in \eqref{eq:Pout_exact_fin} is exact but requires the infinite summations to obtain accurate results where the accuracy depends on the scale and shape parameters of Gamma random variables [\ref{Ref:Moschopoulos}]. Hence, a large number of summations may be necessary for accurate evaluation.

Instead, we find a simpler expression for the outage probability based on the moment-matching approximation. Let a random variable $X$ follow the Gamma distribution $\Gamma(k,\Theta)$. Then the mean and variance of $X$ are given by $\mu = k\Theta$ and $\sigma^2 = k\Theta^2$, respectively. The shape and scale parameters of the Gamma random variable can be expressed with respect to $\mu$ and $\sigma^2$ as $k=\mu^2/\sigma^2$ and $\Theta = \sigma^2/\mu$, respectively. 
As the Gamma random variables $Y_i \sim \Gamma\left(m_i, \frac{a_i}{m_i}\right)$ are independent of one another, their sum, i.e., $\sum_i Y_i$, has the mean of $\mu_{\sum} = \sum_i  \mathbb{E}[Y_i] = \sum_i a_i$ and the variance of $\sigma_{\sum}^2 = \sum_i \mathbf{Var}[Y_i] = \sum_i \frac{a_i^2}{m_i}$.
From this fact, we can find a Gamma random variable $Z \sim \Gamma(\bar{k}, \bar{\Theta})$ with the same first- and second-order moments as $\sum_i Y_i$ where
\begin{align}
\bar{k} = \frac{\mu_{\sum}^2}{\sigma_{\sum}^2} = \frac{(\sum_i a_i)^2}{\sum_i a_i^2/m_i}
\end{align}
and
\begin{align}
\bar{\Theta} = \frac{\sigma_{\sum}^2}{\mu_{\sum}} = \frac{\sum_i a_i^2/m_i}{\sum_i a_i}.
\end{align}
With this moment-matched distribution, the approximated outage probability can be simply obtained as
\begin{align}\label{eq:Pout_approx_fin}
\Pout \approx \Pout^{\mathrm{MM}} = \mathbb{P}[Z<\gth] = \frac{\g(\bar{k},\gth/\bar{\Theta})}{\Gamma(\bar{k})}.
\end{align}

Now we look into the performance behavior in the high SNR regime and obtain the diversity order. Let the power of satellites $P_{\L}=P_{\F_1}=\cdots=P_{\Fn}=P\rightarrow \infty$. Then $\bar{\Theta} \rightarrow \infty$. From the fact that $\lim_{x \rightarrow 0}\g(a,x) \approx x^a/a$ [\ref{Ref:Ding}],
the outage probability in \eqref{eq:Pout_approx_fin} asymptotically becomes 
\begin{align}\label{eq:Pout_asympt_fin}
\Pout^{\mathrm{HS}} = \lim_{P\rightarrow\infty}\Pout^{\mathrm{MM}} \approx \frac{(\gth/\bar{\Theta})^{\bar{k}}}{\bar{k}\Gamma(\bar{k})}.
\end{align}
To analyze the diversity order, we let $a_i=P \bar{a}_i$. 
Then as $P\rightarrow\infty$, $\bar{k}$ converges as
\begin{align} \label{eq:lima}
\bar{k}^{\infty} = \lim_{P\rightarrow\infty}\bar{k}=\lim_{P\rightarrow\infty}\frac{(P\sum_i \bar{a}_i)^2}{P^2\sum_i \bar{a}_i^2/m_i} = \frac{(\sum_i \bar{a}_i)^2}{\sum_i \bar{a}_i^2/m_i}.
\end{align}
The diversity order is then given by
\begin{align}\label{eq:DO}
\mathrm{DO} = -\lim_{P \rightarrow \infty} \frac{\log \Pout^{\mathrm{HS}}}{\log P}
    \mathop=^{(a)}  \lim_{P \rightarrow \infty} \frac{ \bar{k} \log \bar{\Theta}}{\log P} \mathop=^{(b)} \bar{k}^{\infty}
\end{align}
where ($a$) follows from the fact that $\frac{\bar{k}\log\gamma_{\mathrm{th}}}{\log{P}}\rightarrow 0$ as $P\rightarrow \infty$, 
and ($b$) follows from 
\begin{align}\label{eq:logTheta}
\log \bar{\Theta} = \log \!\left(  P^2 \! \sum_i \frac{\bar{a}_i^2}{m_i}\right) \!- \log \left(P \!\sum_i \bar{a}_i\right) = \log P + \mathrm{const}.
\end{align}
\begin{rem}
The diversity order given in \eqref{eq:DO} (or \eqref{eq:lima}) depends on the transceiver and channel parameters. If these parameters are identical for all satellites, the diversity order reduces to $\frac{m(\sum_i d_i^{-\alpha})^2}{\sum_i d_i^{-2\alpha}}$, which is a function of the Nakagami parameter $m$ and the distances $d_i$. This implies the importance of the distance characterization given in Section~\ref{Sec:dist_ch}.    
\end{rem}

\begin{figure}
\begin{center}
\includegraphics[width=\columnwidth]{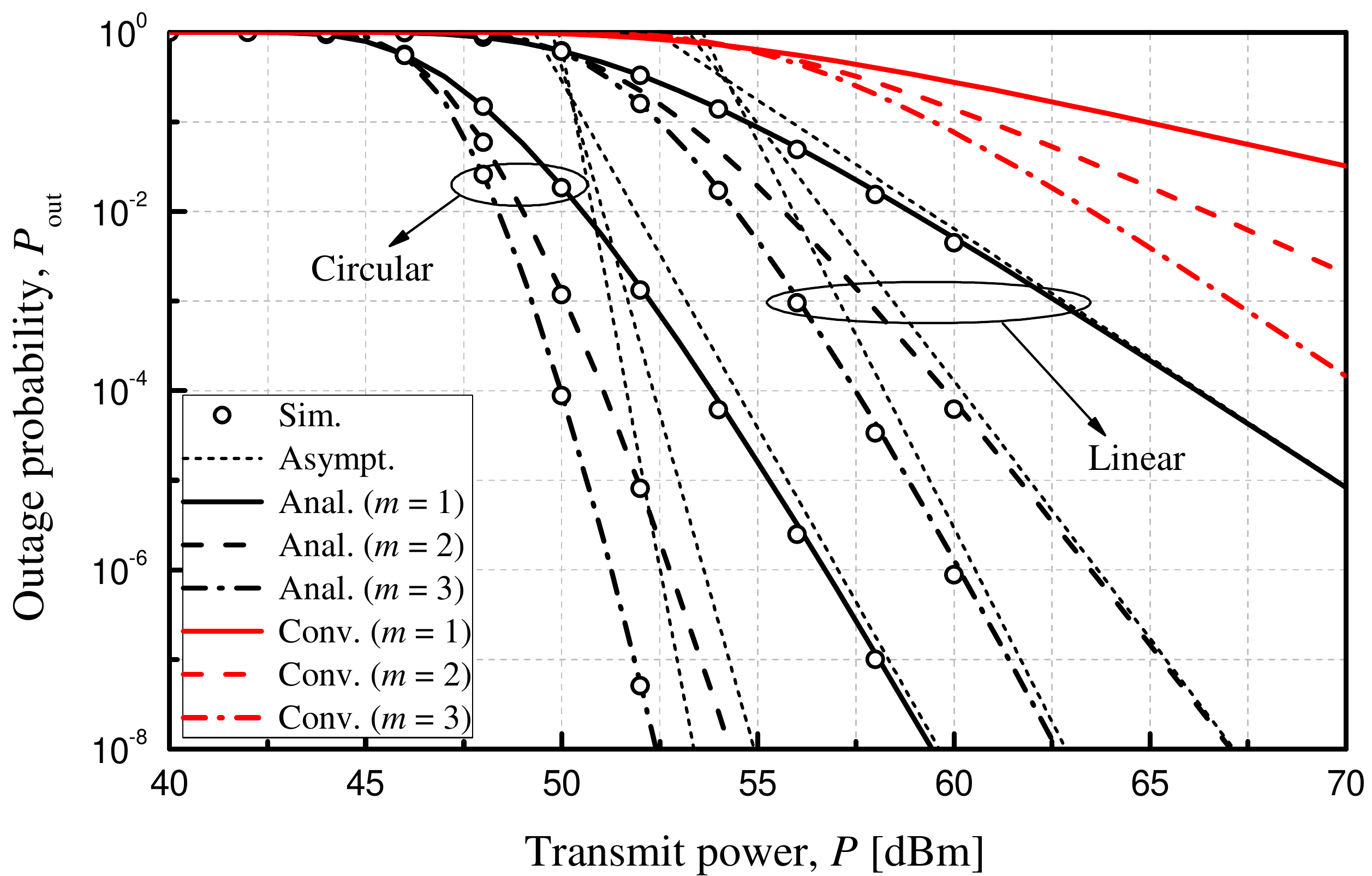}
\end{center}
\setlength\abovecaptionskip{.25ex plus .125ex minus .125ex}
\setlength\belowcaptionskip{.25ex plus .125ex minus .125ex}
\caption{Outage probability versus the transmit power for various Nakagami parameters with $a=600$ km, $i=90$ deg, $u=30$ deg, and $\Omega=130$ deg.}
\vspace{-10pt}
\label{Fig:Pout_vs_P2}
\end{figure}

\section{Simulation Results}\label{Sec:Sim_results}
In this section, we numerically verify the derived results with the following simulation parameters: 
the Earth's radius $\re = 6378$ km, 
the speed of light $c=3\times10^5$ km/s, 
the noise spectral density $N_0 = -174$ dBm/Hz, 
the carrier frequency $f_c=2$ GHz, 
the path loss exponent $\alpha_i=3$,
the transmit and receive antenna gains $\{G_{\mathrm{t},i},G_{\mathrm{r}}\}=\{10,0\}$ dBi, 
and the bandwidth $W=1$ MHz. 
We suppose that the terminal is located in Seoul, South Korea, whose latitude and longitude are 37 and 127 degrees, respectively.
In addition, we set 
the angle offset to the first follower $\beta_0=0$ deg, 
the cluster size $\tc=1$ deg, 
the minimum distance between satellites $d_0=100$ km, 
and the SNR threshold $\gth=-10$ dB.
With this setting, the maximum numbers of followers in the linear and circular clusters are $\Nl = 2$ and $\Nc = 7$.
The analytical results come from the approximated expression in \eqref{eq:Pout_approx_fin}, and the asymptotes are from the asymptotic expression in \eqref{eq:Pout_asympt_fin}.

Fig. \ref{Fig:Pout_vs_P2} shows the outage probability versus the transmit power of the satellites in the cluster for various Nakagami parameters.
Our analysis is in good agreement with the simulation results, and the asymptotes well describe the asymptotic behavior in the high SNR regime.
The satellite clusters have superior performance than conventional systems (denoted as Conv. in the figure) where only a single satellite serves the terminal, i.e., $N=0$. Deploying additional satellites in the conventional systems could also enhance performance, but the enhancement has inherent limitations due to severe inter-satellite interference. Thus, utilizing satellite clusters would be a highly effective solution when numerous satellites will be launched in the future [\ref{Ref:Jung}].
The outage probability decreases with the transmit power of satellites $P$ because the higher transmit power boosts up the SNR at the terminal.
As the Nakagami parameter $m$ increases, the outage performance is improved because a higher value of $m$ indicates less severe fading over the satellite-terminal channels.
It is also shown that the circular clusters outperform the linear clusters because the circular formation has a better spatial efficiency than linear formation given the minimum distance between satellites. 

\begin{figure}
\begin{center}
\includegraphics[width=.87\columnwidth]{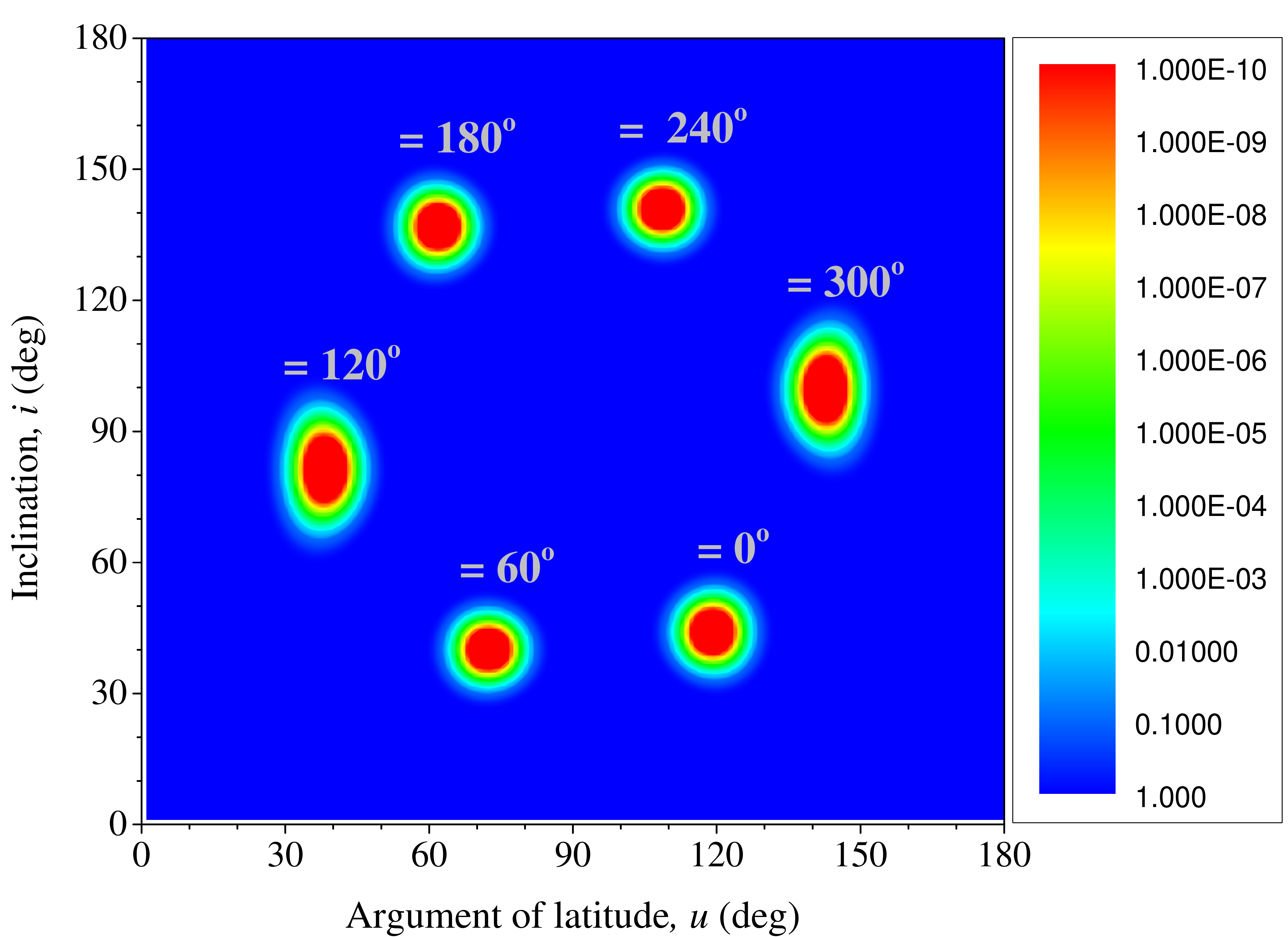}
\end{center}
\setlength\abovecaptionskip{.25ex plus .125ex minus .125ex}
\setlength\belowcaptionskip{.25ex plus .125ex minus .125ex}
\caption{Outage probability versus the inclination and the argument of latitude for various RAANs with $a=600$ km.}
\vspace{-10pt}
\label{Fig:Pout_vs_iu}
\end{figure}

Fig. \ref{Fig:Pout_vs_iu} shows the outage probability depending on the three orbital elements: the inclination $i$, the argument of latitude $u$, and the RAAN $\Omega$. Given any of the three elements, there exist specific ranges of the other elements satisfying a target outage probability. For instance, when $\{\Omega,i\}=\{0, 43\}$ deg, the outage probability less than 0.1 is met within $i\in[109,128]$ deg. From this observation, we can design the orbits of satellite clusters to consistently serve a given area using different combinations of the orbital elements. Additionally, the 180-degree turn in the RAAN makes the ascending and descending nodes flipped so that the outage performance seems symmetric about the point $(u,i)=(90, 90)$ deg.

\vspace{-0.2cm}
\section{Conclusions}\label{Sec:Conclusions}
In this paper, we have effectively analyzed the performance of downlink satellite communication systems with a satellite cluster in a linear or circular formation. By conducting the coordinate transformation and using maximum ratio transmission precoding, we derived the exact expression for the outage probability under the Nakagami fading. We also obtained a simpler approximated expression using second-order moment-matching and analyzed the asymptotic behavior of the outage performance in the high SNR regime. The analytical results were verified through Monte Carlo simulations and are expected to facilitate the performance evaluation of satellite cluster-based communication systems and the design of reliable satellite clusters.

\appendix
\section{}\label{App:Nmax}
\begin{figure}
\begin{center}
\includegraphics[width=0.83\columnwidth]{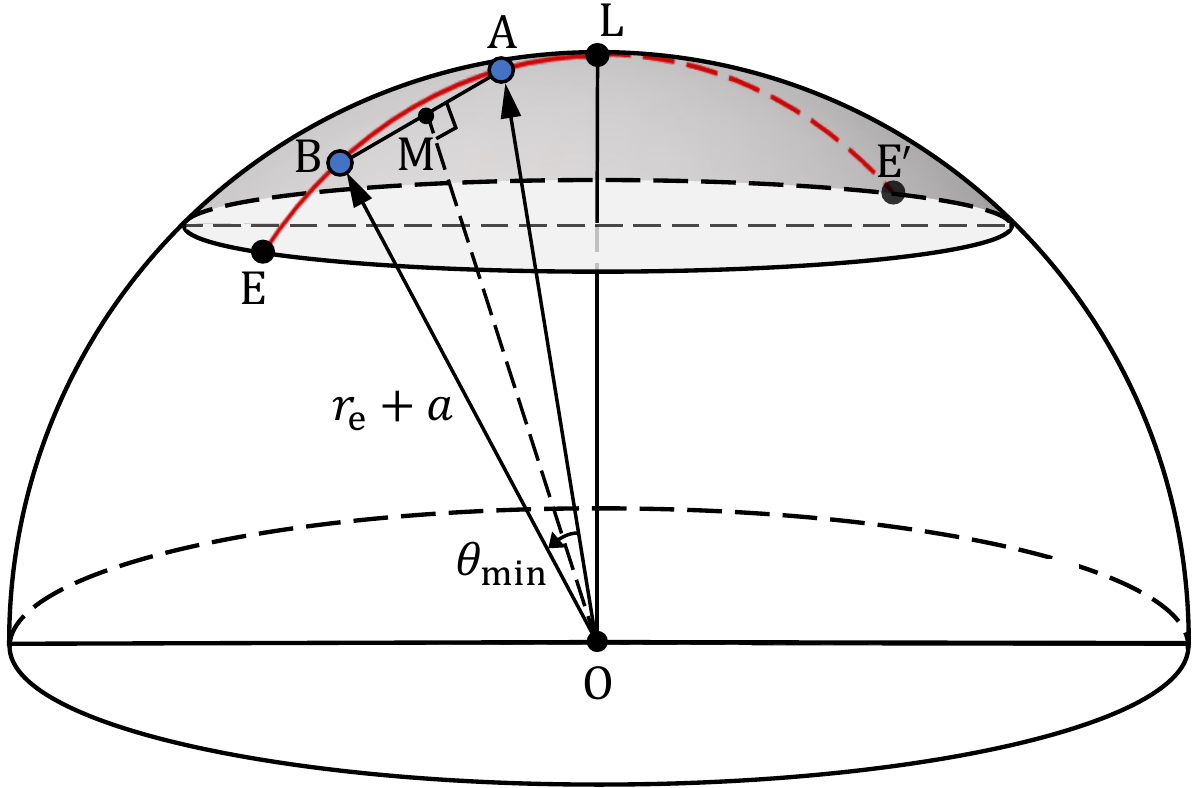}
\end{center}
\setlength\abovecaptionskip{.25ex plus .125ex minus .125ex}
\setlength\belowcaptionskip{.25ex plus .125ex minus .125ex}
\caption{Adjacent two satellites separated by $d_0$ in the linear cluster.}
\vspace{-10pt}
\label{Fig:appA}
\end{figure}

Suppose that two adjacent followers $\mathrm{A}$ and $\mathrm{B}$ in the linear cluster are separated by the minimum distance $d_0$ as in Fig.~\ref{Fig:appA} where $\tmin$ is the minimum angle between $\mathrm{A}$ and $\mathrm{B}$ required to keep the minimum distance.
Let $\mathrm{M}$ denote the midpoint between the two satellites. 
Then, $\tmin$ can be obtained by using the Pythagorean theorem as $\tmin = 2 \angle \mathrm{AOM} = 2 \sin^{-1}\left(\frac{d_0}{2(\re+a)}\right)$,
and the length of the arc between $\mathrm{A}$ and $\mathrm{B}$ is given by $\widehat{\mathrm{AB}}=(\re+a)\tmin$.
We let $\mathrm{E}$ and $\mathrm{E}'$ represent two intersection points between the reference orbit and the circular edge of the spherical cap (shaded area in Fig.  \ref{Fig:appA}). Then, the maximum number of followers is twice the number of followers that can be deployed in $\widehat{\mathrm{EL}}$ i.e., $\Nl = 2 \bigl\lfloor{ \frac{\widehat{\mathrm{EL}}}{\widehat{\mathrm{AB}}} \bigr\rfloor}= 2 \bigl\lfloor{ \frac{\tc}{\tmin} \bigr\rfloor}$.
Similarly, we can obtain the maximum number of followers for the circular cluster by using the circumference of the circular edge, i.e., $2\pi(\re+a)\sin\tc$, as $\Nc 
    = \bigl\lfloor{\frac{2\pi(\re+a)\sin\tc}{\widehat{\mathrm{AB}}}\bigr\rfloor} 
    = \bigl\lfloor{\frac{2\pi\sin\tc}{\tmin}\bigr\rfloor}$.

\ifCLASSOPTIONcaptionsoff
  \newpage
\fi

\end{document}